\documentstyle[preprint,array,aps,epsf]{revtex}

\begin{document}

\draft
%\preprint{}
\title
{Relativistic entanglement of Bell states with general momentum}
\author{Young Hoon Moon$^{1,2}$,Doyeol Ahn$^{1,2}${\footnote{e-mail:dahn@uoscc.uos.ac.kr}},
 and Sung Woo Hwang$^{1,3}${\footnote{e-mail:swhwang@korea.ac.kr}}}
\address{
$^1$Institute of Quantum Information Processing and Systems,
University of Seoul, Seoul, 130-743, Korea\\
$^2$Department of Electrical and Computer Engineering, University
of Seoul,
Seoul, 130-743, Korea\\
$^3$Department of Electronic Engineering, Korea University, Seoul,
136-701, Korea}

\maketitle

\vspace{2.0cm}


\baselineskip24pt

\begin{abstract}
In this paper, the Lorentz transformation of the entangled Bell
states with momentum, not necessarily orthogonal to the boost
direction, and spin, is studied. We extended quantum correlations
and Bell's inequality to the relativistic regime by considering
normalized relativistic observables. It is shown that quantum
information, along the perpendicular direction to the boost, is
eventually lost and Bell's inequality is not always violated for
entangled states in special relativity.  This could impose
restrictions to certain quantum information processing such as
quantum cryptography using massive particles.

\end{abstract}
\vspace{.25in}

%\pacs{PACS numbers: 03.67, 71.10.Li, 71.35.-y, 73.20.D}

%\narrowtext
\newpage
\section{Introduction}
Relativistic quantum information processing is of growing
interest, not only for the logical completeness but also the new
features such as the physical bounds on the information transfers,
processing and the errors provided by the full relativistic
treatment\cite{1}--\cite{11}. It would be also interesting to
study quantum correlations and Bell's inequality  in different
Lorentz frames. Violation of Bell's inequality is perhaps the most
drastic feature distinguishing the quantum theory from the
classical physics\cite{12}. Bell's proof that there are states of
two-quantum-particle systems that do not satisfy the Bell's
inequality derived from Einstein's assumptions\cite{13} of the
principle of local causes has changed our traditional viewpoint of
Nature quite significantly. Specifically, it was shown that all
the non-product states or otherwise known as the entangled states
always violate the Bell inequality when special relativity is not
taken into account\cite{14}.  So it would be an interesting
question to ask if above mentioned condition changes if one
considers special relativity.

Under the Lorentz transformation, the Hilbert space vectors
representing the quantum states undergo the unitary
transformations\cite{15}. On the other hand, the Pauli matrices
are not Lorentz covariant, so there are needs to find
relativistically invariant operators corresponding to the spin in
order to investigate the Bell's inequality within the special
relativity\cite{16}, Sometime ago, Fleming\cite{17} showed that
covariant spin-vector operator which reduces to the ordinary spin
operator in the non-relativistic limit, can be derived from the
Pauli-Lubanski pseudo vector and Czachor\cite{2} showed that the
degree of violation of the Bell's inequality depends on the
velocity  of the pair  of spin$-\frac{1}{2}$ particles with
respect to the laboratory. Unitary transformation corresponding
the Lorentz boost of the quantum states was not considered, in
those works.

In the previous work\cite{6}, we calculated the Bell observables
for entangled states in the rest frame with both momentum vector
and spin in the $z$-direction, seen by the observer moving in the
$x$-direction, and showed that the entangled states do not always
violate the Bell's inequality when the boost speed approaches the
speed of light.  This paper is a direct continuation of a
preceding one\cite{6}(I). In this paper, we study the case of the
general momentum not necessarily in perpendicular to the boost
direction as described in Figures 1 and 2 and derived
transformation rules for the entangled states. We also calculated
the average of the Bell observable for the momentum-conserved
entangled Bell states for spin$-\frac{1}{2}$ particles and show
that Bell's inequality is not always violated for the case of
general momentum in special relativity.  It is also shown that
quantum information, along the perpendicular direction to the
boost, is eventually lost. This could impose restrictions to
certain quantum information processing such as quantum
cryptography using massive particles.  Unless both sender and
receiver measures along the boost direction, there will be
information loss.

\begin{figure}[tb]% \epsffile{fig_1.eps} \epsffile{fig_2.eps}
 \begin{picture}(200,100)(-60,0)
  \put(-62,-62){\makebox(0,0)[tr]{$-\vec{p}$}} \put(63,63){\makebox(0,0)[bl]{$\vec{p}$}}
  \put(-12,-12){\makebox(24,24)[b]{$o$}} \put(228,-12){\makebox(24,24)[b]{$o'$}}
  \put(80,-3){\makebox(12,6)[br]{$x$}} \put(321,-3){\makebox(12,6)[br]{$x'$}}
  \put(-3,80){\makebox(6,12)[t]{$z$}} \put(240,82){\makebox(6,12)[t]{$z'$}}
  \put(160,5){\makebox(0,1)[b]{$\vec{v}$}} \put(10,25){\makebox(0,0){$\theta$}}

  \put(0,0){\vector(1,0){80}} \put(0,0){\vector(0,1){80}}
  \put(120,0){\vector(1,0){80}} \put(240,0){\vector(1,0){80}} \put(240,0){\vector(0,1){80}}
  \qbezier(0,20)(9,20)(14.14,14.14) \put(14.14,14.14){\vector(1,-1){0}} \thicklines
  \put(0,0){\vector(1,1){60}} \put(0,0){\vector(-1,-1){60}}
 \end{picture}
  \centerline{\epsfxsize=10.0cm}
  \vspace*{2.2cm}
  \caption{the case of momentum vector in the $x$-$z$ plane,
$\vec{p}=p(\sin\theta,0,\cos\theta)$ and the boost $\Lambda$ in the $x$-diraction.}
  %\label{fig:SetupFigure}
\end{figure}
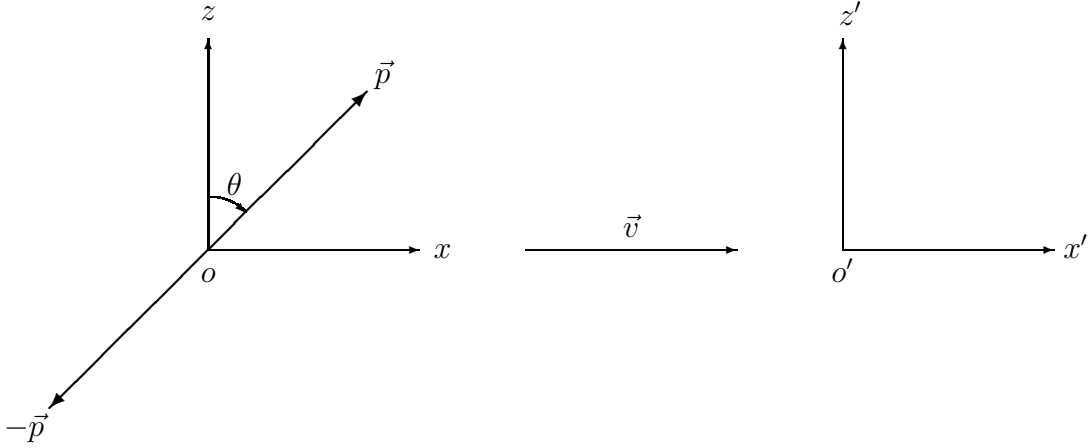

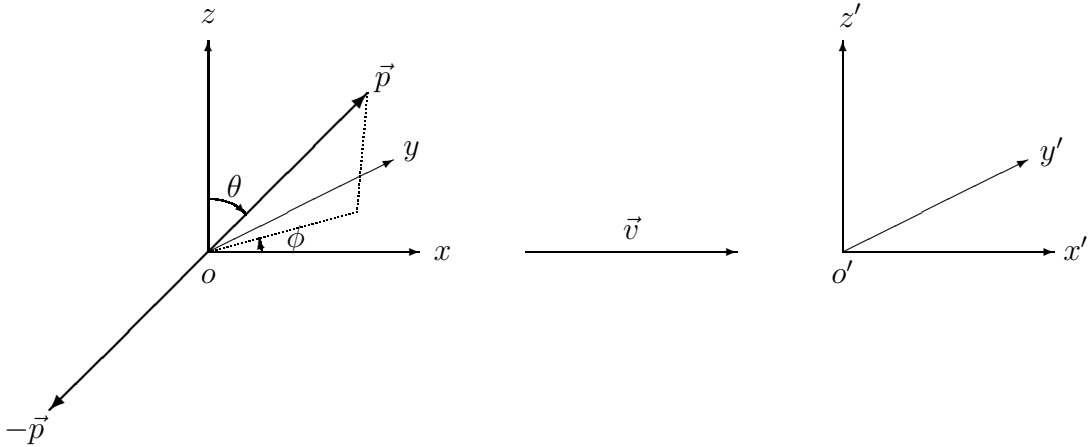
\begin{figure}[t]
 \begin{picture}(100,100)(-60,0)
   \put(-62,-62){\makebox(0,0)[tr]{$-\vec{p}$}} \put(63,63){\makebox(0,0)[bl]{$\vec{p}$}}
   \put(-12,-12){\makebox(24,24)[b]{$o$}} \put(228,-12){\makebox(24,24)[b]{$o'$}}
   \put(80,-3){\makebox(12,6)[br]{$x$}} \put(321,-3){\makebox(12,6)[br]{$x'$}}
   \put(77,38){\makebox(0,0){$y$}} \put(319,38){\makebox(0,0){$y'$}}
   \put(-3,80){\makebox(6,12)[t]{$z$}} \put(240,82){\makebox(6,12)[t]{$z'$}}
   \put(160,5){\makebox(0,1)[b]{$\vec{v}$}} \put(10,25){\makebox(0,0){$\theta$}} \put(33,5){\makebox(0,0){$\phi$}}

   \put(0,0){\vector(1,0){80}} \put(0,0){\vector(2,1){70}} \put(0,0){\vector(0,1){80}}
   \put(120,0){\vector(1,0){80}} \put(240,0){\vector(1,0){80}}
   \put(240,0){\vector(2,1){70}} \put(240,0){\vector(0,1){80}} \qbezier(0,20)(9,20)(14.14,14.14)
   \put(14.14,14.14){\vector(1,-1){0}}
   \qbezier(20,0)(20.1,3)(19,5.2) \put(19,5.2){\vector(-1,4){0}}
   \qbezier[40](0,0)(28,7.5)(56,15) \qbezier[30](60,60)(58,37.5)(56,15) \thicklines
   \put(0,0){\vector(1,1){60}} \put(0,0){\vector(-1,-1){60}}
 \end{picture}
  \centerline{\epsfxsize=10.0cm }
  \vspace*{2.2cm}
  \caption{the case of momentum vector out of plane,
  $\vec{p}=p(\sin\theta\cos\phi,\sin\theta\sin\phi,\cos\theta)$ and the boost in the $x$-direction.}
  %\label{fig:SetupFigure}
\end{figure}

\section{Relativistic entanglements}

A multi-particle state vector is denote by
\begin{equation}
\Psi_{p_1\sigma_1;p_2\sigma_2;\dots}=a^+(\vec{p}_1,
\sigma_1)a^+(\vec{p}_2, \sigma_2)\dots\Psi_0,
\label{states}
\end{equation}
where $p_i$ labels the four-momentum, $\sigma_i$ is the spin $z$ component, $a^+(\vec{p}_i, \sigma_i)$ is the creation operator
which adds a particle with momentum $\vec{p}_i$ and spin $\sigma_i$, and $\Psi_0$ is the Lorentz invariant vacuum state.
The Lorentz transformation $\Lambda$ induces unitary transformation on vectors in the Hilbert space
\begin{equation}
\Psi\rightarrow U(\Lambda)\Psi
\end{equation}
and the operators $U$ satisfies the composition rule
\begin{equation}
U(\bar{\Lambda})U(\Lambda)=U(\bar{\Lambda}\Lambda),
\end{equation}
while the creation operator has the following transformation rule
\begin{equation}
U(\Lambda)a^+(\vec{p},\sigma)U(\Lambda)^{-1}=\sqrt{\frac{(\Lambda p)^0}{p^0}}\sum_{\bar{\sigma}}
{\cal D}^{(j)}_{\bar{\sigma}\sigma}(W(\Lambda, p))a^+(\vec{p}_{\Lambda},\bar{\sigma}).
\label{trans}
\end{equation}
Here, $W(\Lambda, p)$ is the Wigner's little group element given by
\begin{equation}
W(\Lambda, p)=L^{-1}(\Lambda p)\Lambda L(p),
\label{litt}
\end{equation}
with ${\cal D}^{(j)}(W)$ the representation of $W$ for spin $j$, $p^{\mu}=(\vec{p}, p^0)$, $(\Lambda p)^{\mu}=
(\vec{p}_{\Lambda}, (\Lambda p)^0)$ with $\mu=1, 2, 3, 0$ and $L(p)$ is the Lorentz transformation such that
\begin{equation}
p^{\mu}=L^{\mu}_{{}\nu}k^{\nu}
\end{equation}
where $k^{\nu}=(0, 0, 0, m)$ is the four-momentum taken in the particle's rest frame. One can also use the conventional
ket-notation to represent the quantum states as
\begin{eqnarray}
\Psi_{p, \sigma}&=&a^+(\vec{p}, \sigma)\Psi_0\nonumber\\
&=&|\vec{p}, \sigma\rangle\nonumber\\
&=&|\vec{p}\rangle\otimes|\sigma\rangle.
\end{eqnarray}
The Wigner representation of the Lorentz group for the spin-$\frac{1}{2}$ becomes:
\begin{eqnarray}
&&{\cal D}^{(1/2)}(W(\Lambda, p))\nonumber\\
&&=\frac{1}{[(p^0 +m)((\Lambda p)^0 +m)]^{1/2}}\{(p^0 +m)\cosh\frac{\alpha}{2} +
(\vec{p}\cdot\hat{e})\sinh\frac{\alpha}{2}-i \sinh\frac{\alpha}{2}\vec{\sigma}\cdot(\vec{p}\times\hat{e})\}\nonumber\\
&&=\cos\frac{\Omega_{\vec{p}}}{2}+i\sin\frac{\Omega_{\vec{p}}}{2}(\vec{\sigma}\cdot\hat{n}),
\label{rota}
\end{eqnarray}
with
\begin{equation}
\cos\frac{\Omega_{\vec{p}}}{2}=\frac{\cosh\frac{\alpha}{2}\cosh\frac{\delta}{2}+\sinh\frac{\alpha}{2}\sinh\frac{\delta}{2}(\hat{e}\cdot\hat{p})}
{[\frac{1}{2}+\frac{1}{2}\cosh\alpha\cosh\delta+\frac{1}{2}\sinh\alpha\sinh\delta(\hat{e}\cdot\hat{p})]^{1/2}},\label{rotai}
\end{equation}
and
\begin{equation}
\sin\frac{\Omega_{\vec{p}}}{2} \hat{n}=\frac{\sinh\frac{\alpha}{2}\sinh\frac{\delta}{2}(\hat{e}\times\hat{p})}
{[\frac{1}{2}+\frac{1}{2}\cosh\alpha\cosh\delta+\frac{1}{2}\sinh\alpha\sinh\delta(\hat{e}\cdot\hat{p})]^{1/2}},\label{rotaii}
\end{equation}
where $\cosh\delta=\frac{p^0}{m}$.
We note that the eq. (\ref{rota}) indicates the Lorentz group can be represented by the pure rotation by axis $\hat{n}=\hat{e}\times\hat{p}$
for the two-component spinor.

We define the momentum-conserved entangled Bell stats for spin-$\frac{1}{2}$ particles in the rest frame as follows:
\begin{mathletters}
\begin{eqnarray}
\Psi_{00}&=&\frac{1}{\sqrt{2}}\{
a^+(\vec{p},{{1}\over {2}})a^+(-\vec{p},\frac{1}{2})+a^+(\vec{p},-\frac{1}{2})a^+(-\vec{p},-\frac{1}{2})\}\Psi_0,\label{bello}\\
\Psi_{01}&=&\frac{1}{\sqrt{2}}\{
a^+(\vec{p},\frac{1}{2})a^+(-\vec{p},\frac{1}{2})-a^+(\vec{p},-\frac{1}{2})a^+(-\vec{p},-\frac{1}{2})\}\Psi_0,\label{bellt}\\
\Psi_{10}&=&\frac{1}{\sqrt{2}}\{
a^+(\vec{p},\frac{1}{2})a^+(-\vec{p},-\frac{1}{2})+a^+(\vec{p},-\frac{1}{2})a^+(-\vec{p},\frac{1}{2})\}\Psi_0,\label{bellth}\\
\Psi^{11}&=&\frac{1}{\sqrt{2}}\{
a^+(\vec{p},\frac{1}{2})a^+(-\vec{p},-\frac{1}{2})-a^+(\vec{p},-\frac{1}{2})a^+(-\vec{p},\frac{1}{2})\}\Psi_0,\label{bellf}\\
\nonumber
\end{eqnarray}
\end{mathletters}
where $\Psi_0$ is the Lorentz invariant vacuum state.

For an observer in another reference frame $S'$ described by an
arbitrary boost $\Lambda$, the transformed Bell states are given
by
\begin{equation}
\Psi_{ij}\rightarrow U(\Lambda)\Psi_{ij}. \label{ptrans}
\end{equation}

For example, from equations (\ref{trans}) and (\ref{bello}),
$U(\Lambda)\Psi_{00}$ becomes
\begin{eqnarray}
U(\Lambda)\Psi_{00}&=&\frac{1}{\sqrt{2}}\{
U(\Lambda)a^+(\vec{p},\frac{1}{2})U^{-1}(\Lambda)U(\Lambda)a^+(-\vec{p},\frac{1}{2})U^{-1}(\Lambda)\nonumber\\
&&+
U(\Lambda)a^+(\vec{p},-\frac{1}{2})U^{-1}(\Lambda)U(\Lambda)a^+(-\vec{p},-\frac{1}{2})U^{-1}(\Lambda)\}U(\Lambda)\Psi_0
\nonumber\\
&=&\frac{1}{\sqrt{2}}\sum_{\sigma,\sigma'}\{\sqrt{\frac{(\Lambda
p)^0}{p^0}}{\cal D}^{(\frac{1}{2})}_{\sigma\frac{1}{2}}(W(\Lambda,
p))\sqrt{\frac{(\Lambda {\cal P}p)^0}{({\cal P}p)^0}}{\cal
D}^{(\frac{1}{2})}_{\sigma'\frac{1}{2}}(W(\Lambda, {\cal
P}p))a^+(\vec{p}_{\Lambda},\sigma)a^+(-\vec{p}_{\Lambda},\sigma')\nonumber\\
&&+\sqrt{\frac{(\Lambda p)^0}{p^0}}{\cal
D}^{(\frac{1}{2})}_{\sigma-\frac{1}{2}}(W(\Lambda,
p))\sqrt{\frac{(\Lambda {\cal P}p)^0}{({\cal P}p)^0}}{\cal
D}^{(\frac{1}{2})}_{\sigma'-\frac{1}{2}}(W(\Lambda, {\cal
P}p))a^+(\vec{p}_{\Lambda},\sigma)a^+(-\vec{p}_{\Lambda},\sigma')\}\Psi_0
\label{tbell}
\end{eqnarray}
and so on.\\

\textbf{A: The momentum and the boost vectors in the same plane.}

We assume that $\vec{p}$ is in the $x$-$z$ plane,
$\vec{p}=(p\sin\theta,0,p\cos\theta)$ and the boost $\Lambda$ is in $x$-diraction.
In this case, we have
\begin{eqnarray}
\cos\frac{\Omega_{\pm\vec{p}}}{2}&=&\frac{\cosh\frac{\alpha}{2}\cosh\frac{\delta}{2}\pm\sinh\frac{\alpha}{2}\sinh\frac{\delta}{2}\sin\theta}{[\frac{1}{2}+\frac{1}{2}\cosh\alpha\cosh\delta\pm\frac{1}{2}\sinh\alpha\sinh\delta\sin\theta]^{1/2}},\label{cos}\\
\sin\frac{\Omega_{\pm\vec{p}}}{2} \hat{n}_\pm&=&\frac{(\mp\hat{y})\sinh\frac{\alpha}{2}\sinh\frac{\delta}{2}\cos\theta}{[\frac{1}{2}+\frac{1}{2}\cosh\alpha\cosh\delta\pm\frac{1}{2}\sinh\alpha\sinh\delta\sin\theta]^{1/2}},\label{sin}
\end{eqnarray}
and
\begin{eqnarray}
{\cal D}^{1/2}(W(\Lambda, p))&=&\cos\frac{\Omega_{\vec{p}}}{2}-i\sigma_y\sin\frac{\Omega_{\vec{p}}}{2}\nonumber\\
&=&\left( \begin{array}{cc}
          \cos\frac{\Omega_{\vec{p}}}{2} & -\sin\frac{\Omega_{\vec{p}}}{2}\\
          \sin\frac{\Omega_{\vec{p}}}{2} & \cos\frac{\Omega_{\vec{p}}}{2}
          \end{array} \right),\label{wignn}\\
{\cal D}^{1/2}(W(\Lambda, {\cal P}p))&=&\cos\frac{\Omega_{-\vec{p}}}{2}+i\sigma_y\sin\frac{\Omega_{-\vec{p}}}{2}\nonumber\\
&=&\left( \begin{array}{cc}
          \cos\frac{\Omega_{-\vec{p}}}{2} & \sin\frac{\Omega_{-\vec{p}}}{2}\\
          -\sin\frac{\Omega_{-\vec{p}}}{2} & \cos\frac{\Omega_{-\vec{p}}}{2}
          \end{array} \right),\label{wignm}
\end{eqnarray}
where $\hat{n}_\pm=\mp\hat{y}$.

Then from equations (\ref{wignn}),(\ref{wignm}) and (\ref{tbell}), we obtain
\begin{mathletters}
\begin{eqnarray}
U(\Lambda)\Psi_{00}&=&\frac{(\Lambda p)^0}{p^0}\cos\frac{\Omega_{\vec{p}}+\Omega_{-\vec{p}}}{2}
\frac{1}{\sqrt{2}}\{ a^+(\vec{p}_{\Lambda},\frac{1}{2})a^+(-\vec{p}_{\Lambda},\frac{1}{2})+a^+(\vec{p}_{\Lambda},-\frac{1}{2})a^+(-\vec{p}_{\Lambda},-\frac{1}{2})\}\Psi_0\nonumber\\
&&-\frac{(\Lambda p)^0}{p^0}\sin\frac{\Omega_{\vec{p}}+\Omega_{-\vec{p}}}{2}\frac{1}{\sqrt{2}}\{
a^+(\vec{p}_{\Lambda},\frac{1}{2})a^+(-\vec{p}_{\Lambda},-\frac{1}{2})-a^+(\vec{p}_{\Lambda},-\frac{1}{2})a^+(-\vec{p}_{\Lambda},\frac{1}{2})\}\Psi_0\nonumber\\
&=&\frac{(\Lambda p)^0}{p^0}|\vec{p}_{\Lambda}, -\vec{p}_{\Lambda}\rangle\otimes\{\cos\frac{\Omega_{\vec{p}}+\Omega_{-\vec{p}}}{2}\frac{1}{\sqrt{2}}(|\frac{1}{2},\frac{1}{2}\rangle+
|-\frac{1}{2}, -\frac{1}{2}\rangle)\}\nonumber\\
&&-\frac{(\Lambda p)^0}{p^0}|\vec{p}_{\Lambda}, -\vec{p}_{\Lambda}\rangle\otimes\{\sin\frac{\Omega_{\vec{p}}+\Omega_{-\vec{p}}}{2}\frac{1}{\sqrt{2}}(|\frac{1}{2},-\frac{1}{2}\rangle-
|-\frac{1}{2}, \frac{1}{2}\rangle)\}\nonumber\\
&=&\frac{(\Lambda p)^0}{p^0}\{\cos\frac{\Omega_{\vec{p}}+\Omega_{-\vec{p}}}{2} \Psi'_{00} -
\sin\frac{\Omega_{\vec{p}}+\Omega_{-\vec{p}}}{2} \Psi'_{11}\},\label{ulpi}
\end{eqnarray}
where $\Psi'{ij}$ is the Bell states in the moving frame $S'$
whose momenta are transformed as $\vec{p}\to\vec{p}_{\Lambda},-\vec{p}\to-\vec{p}_{\Lambda}$.

Likewise, we have
\begin{eqnarray}
U(\Lambda)\Psi_{01}&=&\frac{(\Lambda p)^0}{p^0}\cos\frac{\Omega_{\vec{p}}-\Omega_{-\vec{p}}}{2}
\frac{1}{\sqrt{2}}\{ a^+(\vec{p}_{\Lambda},\frac{1}{2})a^+(-\vec{p}_{\Lambda},\frac{1}{2})-a^+(\vec{p}_{\Lambda},-\frac{1}{2})a^+(-\vec{p}_{\Lambda},-\frac{1}{2})\}\Psi_0\nonumber\\
&&+\frac{(\Lambda p)^0}{p^0}\sin\frac{\Omega_{\vec{p}}-\Omega_{-\vec{p}}}{2}\frac{1}{\sqrt{2}}\{
a^+(\vec{p}_{\Lambda},\frac{1}{2})a^+(-\vec{p}_{\Lambda},-\frac{1}{2})+a^+(\vec{p}_{\Lambda},-\frac{1}{2})a^+(-\vec{p}_{\Lambda},\frac{1}{2})\}\Psi_0\nonumber\\
&=&\frac{(\Lambda p)^0}{p^0}|\vec{p}_{\Lambda}, -\vec{p}_{\Lambda}\rangle\otimes\{\cos\frac{\Omega_{\vec{p}}-\Omega_{-\vec{p}}}{2}\frac{1}{\sqrt{2}}(|\frac{1}{2},\frac{1}{2}\rangle-
|-\frac{1}{2}, -\frac{1}{2}\rangle)\}\nonumber\\
&&+\frac{(\Lambda p)^0}{p^0}|\vec{p}_{\Lambda}, -\vec{p}_{\Lambda}\rangle\otimes\{\sin\frac{\Omega_{\vec{p}}-\Omega_{-\vec{p}}}{2}\frac{1}{\sqrt{2}}(|\frac{1}{2},-\frac{1}{2}\rangle+
|-\frac{1}{2}, \frac{1}{2}\rangle)\}\nonumber\\
&=&\frac{(\Lambda p)^0}{p^0}\{ \cos\frac{\Omega_{\vec{p}}-\Omega_{-\vec{p}}}{2} \Psi'_{01} +
\sin\frac{\Omega_{\vec{p}}-\Omega_{-\vec{p}}}{2} \Psi'_{10}\},\label{ulpii}
\end{eqnarray}
\begin{eqnarray}
U(\Lambda)\Psi_{10}&=&\frac{(\Lambda p)^0}{p^0}\cos\frac{\Omega_{\vec{p}}-\Omega_{-\vec{p}}}{2}
\frac{1}{\sqrt{2}}\{ a^+(\vec{p}_{\Lambda},\frac{1}{2})a^+(-\vec{p}_{\Lambda},-\frac{1}{2})+a^+(\vec{p}_{\Lambda},-\frac{1}{2})a^+(-\vec{p}_{\Lambda},\frac{1}{2})\}\Psi_0\nonumber\\
&&-\frac{(\Lambda p)^0}{p^0}\sin\frac{\Omega_{\vec{p}}-\Omega_{-\vec{p}}}{2}\frac{1}{\sqrt{2}}\{
a^+(\vec{p}_{\Lambda},\frac{1}{2})a^+(-\vec{p}_{\Lambda},\frac{1}{2})-a^+(\vec{p}_{\Lambda},-\frac{1}{2})a^+(-\vec{p}_{\Lambda},-\frac{1}{2})\}\Psi_0\nonumber\\
&=&\frac{(\Lambda p)^0}{p^0}|\vec{p}_{\Lambda}, -\vec{p}_{\Lambda}\rangle\otimes\{\cos\frac{\Omega_{\vec{p}}-\Omega_{-\vec{p}}}{2}\frac{1}{\sqrt{2}}(|\frac{1}{2},-\frac{1}{2}\rangle+
|-\frac{1}{2}, \frac{1}{2}\rangle)\}\nonumber\\
&&-\frac{(\Lambda p)^0}{p^0}|\vec{p}_{\Lambda}, -\vec{p}_{\Lambda}\rangle\otimes\{\sin\frac{\Omega_{\vec{p}}-\Omega_{-\vec{p}}}{2}\frac{1}{\sqrt{2}}(|\frac{1}{2},\frac{1}{2}\rangle-
|-\frac{1}{2}, -\frac{1}{2}\rangle)\}\nonumber\\&=&\frac{(\Lambda p)^0}{p^0}\{\cos\frac{\Omega_{\vec{p}}-\Omega_{-\vec{p}}}{2} \Psi'_{10} -
\sin\frac{\Omega_{\vec{p}}-\Omega_{-\vec{p}}}{2} \Psi'_{01}\},\label{ulpiii}
\end{eqnarray}
and
\begin{eqnarray}
U(\Lambda)\Psi_{11}&=&\frac{(\Lambda p)^0}{p^0}\cos\frac{\Omega_{\vec{p}}+\Omega_{-\vec{p}}}{2}
\frac{1}{\sqrt{2}}\{ a^+(\vec{p}_{\Lambda},\frac{1}{2})a^+(-\vec{p}_{\Lambda},-\frac{1}{2})-a^+(\vec{p}_{\Lambda},-\frac{1}{2})a^+(-\vec{p}_{\Lambda},\frac{1}{2})\}\Psi_0\nonumber\\
&&+\frac{(\Lambda p)^0}{p^0}\sin\frac{\Omega_{\vec{p}}+\Omega_{-\vec{p}}}{2}\frac{1}{\sqrt{2}}\{
a^+(\vec{p}_{\Lambda},\frac{1}{2})a^+(-\vec{p}_{\Lambda},\frac{1}{2})+a^+(\vec{p}_{\Lambda},-\frac{1}{2})a^+(-\vec{p}_{\Lambda},-\frac{1}{2})\}\Psi_0\nonumber\\
&=&\frac{(\Lambda p)^0}{p^0}|\vec{p}_{\Lambda}, -\vec{p}_{\Lambda}\rangle\otimes\{\cos\frac{\Omega_{\vec{p}}+\Omega_{-\vec{p}}}{2}\frac{1}{\sqrt{2}}(|\frac{1}{2},-\frac{1}{2}\rangle-
|-\frac{1}{2}, \frac{1}{2}\rangle)\}\nonumber\\
&&+\frac{(\Lambda p)^0}{p^0}|\vec{p}_{\Lambda}, -\vec{p}_{\Lambda}\rangle\otimes\{\sin\frac{\Omega_{\vec{p}}+\Omega_{-\vec{p}}}{2}\frac{1}{\sqrt{2}}(|\frac{1}{2},\frac{1}{2}\rangle+
|-\frac{1}{2}, -\frac{1}{2}\rangle)\}\nonumber\\
&=&\frac{(\Lambda p)^0}{p^0}\{\cos\frac{\Omega_{\vec{p}}+\Omega_{-\vec{p}}}{2} \Psi'_{11} +
\sin\frac{\Omega_{\vec{p}}+\Omega_{-\vec{p}}}{2} \Psi'_{00}\}.\label{ulpiv}
\end{eqnarray}
\end{mathletters}
where
\begin{mathletters}
\begin{eqnarray}
\cos\frac{\Omega_{\vec{p}}+ \Omega_{-\vec{p}}}{2}&=&\cos\frac{\Omega_{\vec{p}}}{2}\cos\frac{\Omega_{-\vec{p}}}{2}-\sin\frac{\Omega_{\vec{p}}}{2}\sin\frac{\Omega_{-\vec{p}}}{2}\nonumber\\
&=&\frac{(\cosh\frac{\alpha}{2}\cosh\frac{\delta}{2})^{2}-(\sinh\frac{\alpha}{2}\sinh\frac{\delta}{2})^{2}}
{[(\frac{1}{2}+\frac{1}{2}\cosh\alpha\cosh\delta)^{2}-(\frac{1}{2}\sinh\alpha\sinh\delta\sin\theta) ^{2} ]^{\frac{1}{2}}},\\
\cos\frac{\Omega_{\vec{p}}- \Omega_{-\vec{p}}}{2}&=&\cos\frac{\Omega_{\vec{p}}}{2}\cos\frac{\Omega_{-\vec{p}}}{2}+\sin\frac{\Omega_{\vec{p}}}{2}\sin\frac{\Omega_{-\vec{p}}}{2}\nonumber\\
&=&\frac{(\cosh\frac{\alpha}{2}\cosh\frac{\delta}{2})^{2}+(\sinh\frac{\alpha}{2}\sinh\frac{\delta}{2})^{2}\cos{2}\theta}
{[(\frac{1}{2}+\frac{1}{2}\cosh\alpha\cosh\delta)^{2}-(\frac{1}{2}\sinh\alpha\sinh\delta\sin\theta) ^{2} ]^{\frac{1}{2}}},\\
\sin\frac{\Omega_{\vec{p}}+ \Omega_{-\vec{p}}}{2}&=&\sin\frac{\Omega_{\vec{p}}}{2}\cos\frac{\Omega_{-\vec{p}}}{2}+\cos\frac{\Omega_{\vec{p}}}{2}\sin\frac{\Omega_{-\vec{p}}}{2}\nonumber\\
&=&\frac{2\cosh\frac{\alpha}{2}\cosh\frac{\delta}{2}\sinh\frac{\alpha}{2}\sinh\frac{\delta}{2}\cos\theta}
{[(\frac{1}{2}+\frac{1}{2}\cosh\alpha\cosh\delta)^{2}-(\frac{1}{2}\sinh\alpha\sinh\delta\sin\theta) ^{2} ]^{\frac{1}{2}}},\\
\sin\frac{\Omega_{\vec{p}}- \Omega_{-\vec{p}}}{2}&=&\sin\frac{\Omega_{\vec{p}}}{2}\cos\frac{\Omega_{-\vec{p}}}{2}-\cos\frac{\Omega_{\vec{p}}}{2}\sin\frac{\Omega_{-\vec{p}}}{2}\nonumber\\
&=&\frac{-(\sinh\frac{\alpha}{2}\sinh\frac{\delta}{2})^{2}\sin{2}\theta}
{[(\frac{1}{2}+\frac{1}{2}\cosh\alpha\cosh\delta)^{2}-(\frac{1}{2}\sinh\alpha\sinh\delta\sin\theta) ^{2} ]^{\frac{1}{2}}}.
\end{eqnarray}
\end{mathletters}\\

\textbf{B: The case of the momentum and the boost vectors not in the same plane.}

We consider the general case of momentum vector out of plane,\\
$\vec{p}=(p\sin\theta\cos\phi,p\sin\theta\sin\phi,p\cos\theta)$  and the boost in the $x$-direction.\\
In this case, we have

\begin{eqnarray}
\cos\frac{\Omega_{\pm\vec{p}}}{2}&=&\frac{\cosh\frac{\alpha}{2}\cosh\frac{\delta}{2}\pm\sinh\frac{\alpha}{2}\sinh\frac{\delta}{2}\sin\theta\cos\phi}{[\frac{1}{2}+\frac{1}{2}\cosh\alpha\cosh\delta\pm\frac{1}{2}\sinh\alpha\sinh\delta\sin\theta\cos\phi]^{1/2}},\label{ome21}\\
\sin\frac{\Omega_{\pm\vec{p}}}{2} \hat{n}_\pm&=&\frac{ r \hat{n}_\pm\sinh\frac{\alpha}{2}\sinh\frac{\delta}{2}}{[\frac{1}{2}+\frac{1}{2}\cosh\alpha\cosh\delta\pm\frac{1}{2}\sinh\alpha\sinh\delta\sin\theta\cos\phi]^{1/2}},\label{ome22}
\end{eqnarray}

and

\begin{eqnarray}
{\cal D}^{1/2}(W(\Lambda, p))&=&\cos\frac{\Omega_{\vec{p}}}{2}+i\sin\frac{\Omega_{\vec{p}}}{2}\vec{\sigma}\cdot(-\hat{y}\cos\eta+\hat{z}\sin\eta)\nonumber\\
&=&\left( \begin{array}{cc}
          \cos\frac{\Omega_{\vec{p}}}{2}+i\sin\frac{\Omega_{\vec{p}}}{2}\sin\eta & -\sin\frac{\Omega_{\vec{p}}}{2}\cos\eta\\
          \sin\frac{\Omega_{\vec{p}}}{2}\cos\eta & \cos\frac{\Omega_{\vec{p}}}{2}-i\sin\frac{\Omega_{\vec{p}}}{2}\sin\eta
          \end{array} \right),\label{wignx}\\
{\cal D}^{1/2}(W(\Lambda, {\cal P}p))&=&\cos\frac{\Omega_{-\vec{p}}}{2}-i\sin\frac{\Omega_{-\vec{p}}}{2}\vec{\sigma}\cdot(-\hat{y}\cos\eta+\hat{z}\sin\eta)\nonumber\\
&=&\left( \begin{array}{cc}
          \cos\frac{\Omega_{-\vec{p}}}{2}-i\sin\frac{\Omega_{-\vec{p}}}{2}\sin\eta & \sin\frac{\Omega_{-\vec{p}}}{2}\cos\eta\\
          -\sin\frac{\Omega_{-\vec{p}}}{2}\cos\eta & \cos\frac{\Omega_{-\vec{p}}}{2}+i\sin\frac{\Omega_{-\vec{p}}}{2}\sin\eta
          \end{array} \right)
\label{wigny}
\end{eqnarray}

where $\hat{n}_\pm=\pm(-\hat{y}\cos\eta+\hat{z}\sin\eta),\cos\eta=\frac{\cos\theta}{r},\sin\eta=\frac{\sin\theta\sin\phi}{r},r=\sqrt{\sin^{2}\theta\sin^{2}\phi +\cos^{2}\theta}$.\\

Let $\bar{\Omega}_{\vec{p}}=\frac{\Omega_{\vec{p}}+\Omega_{-\vec{p}}}{2},\Delta\Omega_{\vec{p}}=\frac{\Omega_{\vec{p}}-\Omega_{-\vec{p}}}{2}$.\\
Then from equations (\ref{wignx}),(\ref{wigny}) and (\ref{tbell}), we obtain
\begin{mathletters}
\begin{eqnarray}
U(\Lambda)\Psi_{00}&=&\frac{(\Lambda p)^0}{p^0}(\cos\bar{\Omega}_{\vec{p}}\cos^{2}\eta+\cos\Delta\Omega_{\vec{p}}\sin^{2}\eta)\frac{1}{\sqrt{2}}\{ a^+(\vec{p}_{\Lambda},\frac{1}{2})a^+(-\vec{p}_{\Lambda},\frac{1}{2})\nonumber\\ &&+a^+(\vec{p}_{\Lambda},-\frac{1}{2})a^+(-\vec{p}_{\Lambda},-\frac{1}{2})\}\Psi_0\nonumber\\
&&-\frac{(\Lambda p)^0}{p^0}\sin\bar{\Omega}_{\vec{p}}\cos\eta\frac{1}{\sqrt{2}}\{
a^+(\vec{p}_{\Lambda},\frac{1}{2})a^+(-\vec{p}_{\Lambda},-\frac{1}{2})-a^+(\vec{p}_{\Lambda},-\frac{1}{2})a^+(-\vec{p}_{\Lambda},\frac{1}{2})\}\Psi_0\nonumber\\
&&+i\frac{(\Lambda p)^0}{p^0}\sin\Delta\Omega_{\vec{p}}\sin\eta\frac{1}{\sqrt{2}}\{
a^+(\vec{p}_{\Lambda},\frac{1}{2})a^+(-\vec{p}_{\Lambda},\frac{1}{2})-a^+(\vec{p}_{\Lambda},-\frac{1}{2})a^+(-\vec{p}_{\Lambda},-\frac{1}{2})\}\Psi_0\nonumber\\
&&-i\frac{(\Lambda p)^0}{p^0}(-\cos\bar{\Omega}_{\vec{p}}+\cos\Delta\Omega_{\vec{p}})\sin\eta\cos\eta\frac{1}{\sqrt{2}}\{
a^+(\vec{p}_{\Lambda},\frac{1}{2})a^+(-\vec{p}_{\Lambda},-\frac{1}{2})\nonumber\\ &&+a^+(\vec{p}_{\Lambda},-\frac{1}{2})a^+(-\vec{p}_{\Lambda},\frac{1}{2})\}\Psi_0\nonumber\\
&=&\frac{(\Lambda p)^0}{p^0}|\vec{p}_{\Lambda}, -\vec{p}_{\Lambda}\rangle\otimes\{(\cos\bar{\Omega}_{\vec{p}}\cos^{2}\eta+\cos\Delta\Omega_{\vec{p}}\sin^{2}\eta)\frac{1}{\sqrt{2}}(|\frac{1}{2},\frac{1}{2}\rangle+
|-\frac{1}{2}, -\frac{1}{2}\rangle)\}\nonumber\\
&&-\frac{(\Lambda p)^0}{p^0}|\vec{p}_{\Lambda}, -\vec{p}_{\Lambda}\rangle\otimes\{\sin\bar{\Omega}_{\vec{p}}\cos\eta\frac{1}{\sqrt{2}}(|\frac{1}{2},-\frac{1}{2}\rangle-
|-\frac{1}{2}, \frac{1}{2}\rangle)\}\nonumber\\
&&+i\frac{(\Lambda p)^0}{p^0}|\vec{p}_{\Lambda}, -\vec{p}_{\Lambda}\rangle\otimes\{\sin\Delta\Omega_{\vec{p}}\sin\eta\frac{1}{\sqrt{2}}(|\frac{1}{2},\frac{1}{2}\rangle-
|-\frac{1}{2}, -\frac{1}{2}\rangle)\}\nonumber\\
&&-i\frac{(\Lambda p)^0}{p^0}|\vec{p}_{\Lambda}, -\vec{p}_{\Lambda}\rangle\otimes\{(-\cos\bar{\Omega}_{\vec{p}}+\cos\Delta\Omega_{\vec{p}})\sin\eta\cos\eta\frac{1}{\sqrt{2}}(|\frac{1}{2},-\frac{1}{2}\rangle+
|-\frac{1}{2}, \frac{1}{2}\rangle)\}\nonumber\\
&=&\frac{(\Lambda p)^0}{p^0}\{(\cos\bar{\Omega}_{\vec{p}}\cos^{2}\eta+\cos\Delta\Omega_{\vec{p}}\sin^{2}\eta) \Psi'_{00} -\sin\bar{\Omega}_{\vec{p}}\cos\eta \Psi'_{11}\nonumber\\
&&+i\sin\Delta\Omega_{\vec{p}}\sin\eta \Psi'_{01}-i(-\cos\bar{\Omega}_{\vec{p}}+\cos\Delta\Omega_{\vec{p}})\sin\eta\cos\eta \Psi'_{10}\},\label{ulti}
\end{eqnarray}
where $\Psi'{ij}$ is the Bell states in the moving frame $S'$
whose momenta are transformed as $\vec{p}\to\vec{p}_{\Lambda},
-\vec{p}\to-\vec{p}_{\Lambda}$. Likewise, we have
\begin{eqnarray}
U(\Lambda)\Psi_{01}&=&\frac{(\Lambda p)^0}{p^0}\cos\Delta\Omega_{\vec{p}}\frac{1}{\sqrt{2}}\{ a^+(\vec{p}_{\Lambda},\frac{1}{2})a^+(-\vec{p}_{\Lambda},\frac{1}{2})-a^+(\vec{p}_{\Lambda},-\frac{1}{2})a^+(-\vec{p}_{\Lambda},-\frac{1}{2})\}\Psi_0\nonumber\\
&&+\frac{(\Lambda p)^0}{p^0}\sin\Delta\Omega_{\vec{p}}\cos\eta\frac{1}{\sqrt{2}}\{
a^+(\vec{p}_{\Lambda},\frac{1}{2})a^+(-\vec{p}_{\Lambda},-\frac{1}{2})+a^+(\vec{p}_{\Lambda},-\frac{1}{2})a^+(-\vec{p}_{\Lambda},\frac{1}{2})\}\Psi_0\nonumber\\
&&+i\frac{(\Lambda p)^0}{p^0}\sin\Delta\Omega_{\vec{p}}\sin\eta\frac{1}{\sqrt{2}}\{
a^+(\vec{p}_{\Lambda},\frac{1}{2})a^+(-\vec{p}_{\Lambda},\frac{1}{2})+a^+(\vec{p}_{\Lambda},-\frac{1}{2})a^+(-\vec{p}_{\Lambda},-\frac{1}{2})\}\Psi_0\nonumber\\
&=&\frac{(\Lambda p)^0}{p^0}|\vec{p}_{\Lambda}, -\vec{p}_{\Lambda}\rangle\otimes\{\cos\Delta\Omega_{\vec{p}}\frac{1}{\sqrt{2}}(|\frac{1}{2},\frac{1}{2}\rangle-
|-\frac{1}{2}, -\frac{1}{2}\rangle)\}\nonumber\\
&&+\frac{(\Lambda p)^0}{p^0}|\vec{p}_{\Lambda}, -\vec{p}_{\Lambda}\rangle\otimes\{\sin\Delta\Omega_{\vec{p}}\cos\eta\frac{1}{\sqrt{2}}(|\frac{1}{2},-\frac{1}{2}\rangle+
|-\frac{1}{2}, \frac{1}{2}\rangle)\}\nonumber\\
&&+i\frac{(\Lambda p)^0}{p^0}|\vec{p}_{\Lambda}, -\vec{p}_{\Lambda}\rangle\otimes\{\sin\Delta\Omega_{\vec{p}}\sin\eta\frac{1}{\sqrt{2}}(|\frac{1}{2},\frac{1}{2}\rangle+
|-\frac{1}{2}, -\frac{1}{2}\rangle)\}\nonumber\\
&=&\frac{(\Lambda p)^0}{p^0}\{\cos\Delta\Omega_{\vec{p}} \Psi'_{01} +
\sin\Delta\Omega_{\vec{p}}\cos\eta \Psi'_{10}+i\sin\Delta\Omega_{\vec{p}}\sin\eta \Psi'_{00}\},\label{ultii}
\end{eqnarray}
\begin{eqnarray}
U(\Lambda)\Psi_{10}&=&\frac{(\Lambda p)^0}{p^0}(\cos\bar{\Omega}_{\vec{p}}\sin^{2}\eta+\cos\Delta\Omega_{\vec{p}}\cos^{2}\eta)\frac{1}{\sqrt{2}}\{
a^+(\vec{p}_{\Lambda},\frac{1}{2})a^+(-\vec{p}_{\Lambda},-\frac{1}{2})\nonumber\\ &&+a^+(\vec{p}_{\Lambda},-\frac{1}{2})a^+(-\vec{p}_{\Lambda},\frac{1}{2})\}\Psi_0\nonumber\\
&&-\frac{(\Lambda p)^0}{p^0}\sin\Delta\Omega_{\vec{p}}\cos\eta\frac{1}{\sqrt{2}}\{
a^+(\vec{p}_{\Lambda},\frac{1}{2})a^+(-\vec{p}_{\Lambda},\frac{1}{2})-a^+(\vec{p}_{\Lambda},-\frac{1}{2})a^+(-\vec{p}_{\Lambda},-\frac{1}{2})\}\Psi_0\nonumber\\
&&+i\frac{(\Lambda p)^0}{p^0}\sin\bar{\Omega}_{\vec{p}}\sin\eta\frac{1}{\sqrt{2}}\{
a^+(\vec{p}_{\Lambda},\frac{1}{2})a^+(-\vec{p}_{\Lambda},-\frac{1}{2})-a^+(\vec{p}_{\Lambda},-\frac{1}{2})a^+(-\vec{p}_{\Lambda},\frac{1}{2})\}\Psi_0\nonumber\\
&&-i\frac{(\Lambda p)^0}{p^0}(\cos\bar{\Omega}_{\vec{p}}-\cos\Delta\Omega_{\vec{p}})\sin\eta\cos\eta\frac{1}{\sqrt{2}}\{
a^+(\vec{p}_{\Lambda},\frac{1}{2})a^+(-\vec{p}_{\Lambda},\frac{1}{2})\nonumber\\ &&+a^+(\vec{p}_{\Lambda},-\frac{1}{2})a^+(-\vec{p}_{\Lambda},-\frac{1}{2})\}\Psi_0\nonumber\\
&=&\frac{(\Lambda p)^0}{p^0}|\vec{p}_{\Lambda}, -\vec{p}_{\Lambda}\rangle\otimes\{(\cos\bar{\Omega}_{\vec{p}}\sin^{2}\eta+\cos\Delta\Omega_{\vec{p}}\cos^{2}\eta)\frac{1}{\sqrt{2}}(|\frac{1}{2},-\frac{1}{2}\rangle+
|-\frac{1}{2}, \frac{1}{2}\rangle)\}\nonumber\\
&&-\frac{(\Lambda p)^0}{p^0}|\vec{p}_{\Lambda}, -\vec{p}_{\Lambda}\rangle\otimes\{\sin\Delta\Omega_{\vec{p}}\cos\eta\frac{1}{\sqrt{2}}(|\frac{1}{2},\frac{1}{2}\rangle-
|-\frac{1}{2}, -\frac{1}{2}\rangle)\}\nonumber\\
&&+i\frac{(\Lambda p)^0}{p^0}|\vec{p}_{\Lambda}, -\vec{p}_{\Lambda}\rangle\otimes\{\sin\bar{\Omega}_{\vec{p}}\sin\eta\frac{1}{\sqrt{2}}(|\frac{1}{2},-\frac{1}{2}\rangle-
|-\frac{1}{2}, \frac{1}{2}\rangle)\}\nonumber\\
&&-i\frac{(\Lambda p)^0}{p^0}|\vec{p}_{\Lambda}, -\vec{p}_{\Lambda}\rangle\otimes\{(\cos\bar{\Omega}_{\vec{p}}-\cos\Delta\Omega_{\vec{p}})\cos\eta\sin\eta\frac{1}{\sqrt{2}}(|\frac{1}{2},\frac{1}{2}\rangle+
|-\frac{1}{2}, -\frac{1}{2}\rangle)\}\nonumber\\
&=&\frac{(\Lambda p)^0}{p^0}\{(\cos\bar{\Omega}_{\vec{p}}\sin^{2}\eta+\cos\Delta\Omega_{\vec{p}}\cos^{2}\eta) \Psi'_{10} -\sin\Delta\Omega_{\vec{p}}\cos\eta \Psi'_{01}\nonumber\\
&&+i\sin\bar{\Omega}_{\vec{p}}\sin\eta \Psi'_{11}-i(\cos\bar{\Omega}_{\vec{p}}-\cos\Delta\Omega_{\vec{p}})\cos\eta\sin\eta \Psi'_{00}\},\label{ultiii}
\end{eqnarray}
and
\begin{eqnarray}
U(\Lambda)\Psi_{11}&=&\frac{(\Lambda p)^0}{p^0}\cos\bar{\Omega}_{\vec{p}}\frac{1}{\sqrt{2}}\{
a^+(\vec{p}_{\Lambda},\frac{1}{2})a^+(-\vec{p}_{\Lambda},-\frac{1}{2})-a^+(\vec{p}_{\Lambda},-\frac{1}{2})a^+(-\vec{p}_{\Lambda},\frac{1}{2})\}\Psi_0\nonumber\\
&&+\frac{(\Lambda p)^0}{p^0}\sin\bar{\Omega}_{\vec{p}}\cos\eta\frac{1}{\sqrt{2}}\{
a^+(\vec{p}_{\Lambda},\frac{1}{2})a^+(-\vec{p}_{\Lambda},\frac{1}{2})+a^+(\vec{p}_{\Lambda},-\frac{1}{2})a^+(-\vec{p}_{\Lambda},-\frac{1}{2})\}\Psi_0\nonumber\\
&&+i\frac{(\Lambda p)^0}{p^0}\sin\bar{\Omega}_{\vec{p}}\sin\eta\frac{1}{\sqrt{2}}\{
a^+(\vec{p}_{\Lambda},\frac{1}{2})a^+(-\vec{p}_{\Lambda},-\frac{1}{2})+a^+(\vec{p}_{\Lambda},-\frac{1}{2})a^+(-\vec{p}_{\Lambda},\frac{1}{2})\}\Psi_0\nonumber\\
&=&\frac{(\Lambda p)^0}{p^0}|\vec{p}_{\Lambda}, -\vec{p}_{\Lambda}\rangle\otimes\{\cos\bar{\Omega}_{\vec{p}}\frac{1}{\sqrt{2}}(|\frac{1}{2},-\frac{1}{2}\rangle-
|-\frac{1}{2}, \frac{1}{2}\rangle)\}\nonumber\\
&&+\frac{(\Lambda p)^0}{p^0}|\vec{p}_{\Lambda}, -\vec{p}_{\Lambda}\rangle\otimes\{\sin\bar{\Omega}_{\vec{p}}\cos\eta\frac{1}{\sqrt{2}}(|\frac{1}{2},\frac{1}{2}\rangle+
|-\frac{1}{2}, -\frac{1}{2}\rangle)\}\nonumber\\
&&+i\frac{(\Lambda p)^0}{p^0}|\vec{p}_{\Lambda}, -\vec{p}_{\Lambda}\rangle\otimes\{\sin\bar{\Omega}_{\vec{p}}\sin\eta\frac{1}{\sqrt{2}}(|\frac{1}{2},-\frac{1}{2}\rangle+
|-\frac{1}{2}, \frac{1}{2}\rangle)\}\nonumber\\
&=&\frac{(\Lambda p)^0}{p^0}\{\cos\bar{\Omega}_{\vec{p}} \Psi'_{11} +
\sin\bar{\Omega}_{\vec{p}}\cos\eta \Psi'_{00}+i\sin\bar{\Omega}_{\vec{p}}\sin\eta \Psi'_{10}\}.\label{ultiv}
\end{eqnarray}
\end{mathletters}

If we regard $\Psi'_{ij}$ as Bell states
in the moving frame $S'$, then to an observer in $S'$, the effects
of the Lorentz transformation on entangled Bell
states among themselves should appear as rotations of Bell states in the
frame $S'$.

\section{Bell's inequality.}

We are now, ready to check whether the Lorentz transformed Bell
states always violate the Bell's inequality in special relativity.

One of the most essential features of quantum mechanics
distinguished from the classical physics is that the expectation
value, or the quantum correlation of the measurement of the
observables $\vec{\alpha}_{1}\cdot\vec{\sigma}_{1}$ and
$\vec{\alpha}_{2}\cdot\vec{\sigma}_{2}$ for two-particle system,
where $\vec{\sigma}_{1}$ and $\vec{\sigma}_{2}$ are the Pauli spin
matrices pertaining to the two particles and $\vec{\alpha}_{1}$
and $\vec{\alpha}_{2}$ are unit vectors, given by\cite{18}
\begin{equation}
\langle\vec{\alpha}_{1}\cdot\vec{\sigma}_{1}\vec{\alpha}_{2}\cdot\vec{\sigma}_{2}\rangle=-\vec{\alpha}_{1}\cdot\vec{\alpha}_{2}
\end{equation}
for the singlet state, is always stronger than the classical
correlations. Original Bell's inequality was derived for any
physical system with dichotomic observables, whose values are
$\pm1$ . Since any Hermitian operator defines an observables,  one
could extend the Bell's inequality to the relativistic regime for
any normalized relativistic observables.

It is already known\cite{16} that neither the rest frame spin
$\vec{\sigma}$ nor the Dirac Spin operator $\vec{\Sigma}$ which is
associated with the spin of a moving particle as seen by a
stationary observer can not be the relativistic spin operator.
Another plausible candidate is the Pauli-Lubanski pseudovector
$\emph{W}~^\mu$ which itself is a Casimir operator satisfying
$\emph{W}~^\mu\emph{W}_\mu=m^{2}s(s+1)$, where $m$ and $s$ are the
mass and spin of the particle, respectively, and
$\emph{W}~^\mu=(p^{0}(\vec{e}\cdot\vec{s})\vec{e}+mc(\vec{s}-(\vec{e}\cdot\vec{s})\vec{e}),p^{0}\vec{v}\cdot\vec{s}/c^{2}
)$ for the observer in the moving frame with boost velocity
$\vec{v}$\cite{1,17,19}. Here $\vec{s}$~ is the spin vector in the
rest frame, $\vec{e}$~ is the unit vector in the Lorentz boost
direction, and $\beta=v/c$ , the ratio of the boost speed and the
speed of light.

In non-relativistic case, the measurement of the spin in the
direction of the unit vector direction $\vec{a}$ is represented by
the observable $\vec{a}\cdot\vec{s}$ and if we extend this
definition of the observable to the relativistic case as
$\vec{a}\cdot\vec{s}_{\Lambda}$, then
\begin{equation}
\vec{a}\cdot\vec{s}_{\Lambda} =[\sqrt{1-\beta^{2}}(\vec{a}-\vec{e}(\vec{a}\cdot\vec{e}))+\vec{e}(\vec{a}\cdot\vec{e})]\cdot\vec{s}
\end{equation}
Here we postulate the relativistic spin as\cite{1,6}
\begin{equation}
\vec{s}_{\Lambda} =
\frac{mc}{p^{0}}\vec{s}+(1-\frac{mc}{p^{0}})(\vec{e}\cdot\vec{s})\vec{e}=\sqrt{1-\beta^{2}}(\vec{s}-\vec{e}(\vec{s}\cdot\vec{e}))+\vec{e}(\vec{s}\cdot\vec{e})=\vec{\emph{W}}/p^{0},
\label{belle1}
\end{equation}
and the normalized relativistic spin observable   is given
by\cite{1,6}
\begin{equation}
\hat{a}=\frac{[\sqrt{1-\beta^{2}}(\vec{a}-\vec{e}(\vec{a}\cdot\vec{e}))+\vec{e}(\vec{a}\cdot\vec{e})]}{\sqrt{1+\beta^2[(\hat{e}\cdot\vec{a})^2-1]}}\cdot\vec{\sigma},
\label{belle2}
\end{equation}
where we normalized the relativisitic spin observable by the
absolute value of its eigenvalue. Here $\vec{a}$ and
$\vec{s}_{\Lambda}$ are unit direction vector and relativistic
spin operator seen by the moving observer.  We can give more clear
physical meaning of Eqs. (\ref{belle1}) and (\ref{belle2}) by
invoking the principle of the special relativity.  If we note
$\vec{a}_{\Lambda}$ as the Lorentz  transformation (now seen in
the rest frame) of direction vector   (of the moving frame), then
from eqs. (\ref{belle1}) and (\ref{belle2}), we obtain
\begin{equation}
\frac{\vec{a}_{\Lambda}\cdot\vec{s}}{|\lambda(\vec{a}_{\Lambda}\cdot\vec{s})|}=\frac{\vec{a}\cdot\vec{s}_{\Lambda}}{|\lambda(\vec{a}\cdot\vec{s}_{\Lambda})|},
\end{equation}
which is consistent with the principle of the special relativity which tells the physics doesn't change
across the frame. As a result we can interpret $\hat{a}$ as correct normalized relativistic
observable for the observer in the moving frame.  Here $\lambda(\hat{O})$  denotes the eigenvalue of
an operator $\hat{O}$.

It is straightforward to calculate the classical correlation
$\langle\hat{a}\hat{b}\rangle_{\textit{classical}}$ when the
moving observer is receding (approaching) from (to) the rest frame
with the speed of light and is given by
\begin{equation}
\langle\hat{a}\hat{b}\rangle_{\textit{classical}}=\frac{\vec{a}\cdot\vec{e}}{|\vec{a}\cdot\vec{e}|}\cdot\frac{\vec{b}\cdot\vec{e}}{|\vec{b}\cdot\vec{e}|}=\pm1,
\end{equation}
and it should be noted that the information in the perpendicular direction to the unit boost vector $\vec{e}$
is lost as both spins are titled toward the boost axis as a result of the Lorentz  transformation.

Normalized relativistic spin observables $\hat{a}, \hat{b}$ are
given by \cite{6}
\begin{equation}
\hat{a}=\frac{(\sqrt{1-\beta^2} \vec{a}_{\perp}+\vec{a}_{\parallel})\cdot\vec{\sigma}}{\sqrt{1+\beta^2[(\hat{e}\cdot\vec{a})^2-1]}}
\label{hata}
\end{equation}
and
\begin{equation}
\hat{b}=\frac{(\sqrt{1-\beta^2} \vec{b}_{\perp}+\vec{b}_{\parallel})\cdot\vec{\sigma}}{\sqrt{1+\beta^2[(\hat{e}\cdot\vec{b})^2-1]}},
\label{hatb}
\end{equation}
where the subscript $\perp$ and $\parallel$ denote the components of
$\vec{a}$ (or $\vec{b}$) which are perpendicular and parallel to the boost direction, respectively. Moreover, $|\vec{a}|=|\vec{b}|=1$.\\

\textbf{A: The momentum and the boost vectors in the same plane.}

Case I: $\Psi_{00}\rightarrow U(\Lambda)\Psi_{00}$\\
From eq. (\ref{ulpi}), we have
\begin{eqnarray}
U(\Lambda)\Psi_{00}&=&\frac{(\Lambda p)^0}{p^0}|\vec{p}_{\Lambda}, -\vec{p}_{\Lambda}\rangle\otimes
[\frac{1}{\sqrt{2}}\cos\frac{\Omega_{\vec{p}}+\Omega_{-\vec{p}}}{2}(|\frac{1}{2},\frac{1}{2}\rangle+|-\frac{1}{2}, -\frac{1}{2}\rangle)\nonumber\\
&&-\frac{1}{\sqrt{2}}\sin\frac{\Omega_{\vec{p}}+\Omega_{-\vec{p}}}{2}(|\frac{1}{2},-\frac{1}{2}\rangle-|-\frac{1}{2}, \frac{1}{2}\rangle)].
\end{eqnarray}
Then, after some mathematical manipulations, we get
\begin{mathletters}
\begin{eqnarray}
\hat{a}\otimes\hat{b}|\frac{1}{2},\frac{1}{2}\rangle&=&\frac{1}{\sqrt{[1+\beta^2(a_x^2-1)][1+\beta^2(b_x^2-1)]}}\{(1-\beta^2)a_z b_z|\frac{1}{2},\frac{1}{2}\rangle\nonumber\\
&&+\sqrt{1-\beta^2}a_z (b_x+ib_y\sqrt{1-\beta^2})|\frac{1}{2},-\frac{1}{2}\rangle\nonumber\\
&&+\sqrt{1-\beta^2}b_z (a_x+ia_y\sqrt{1-\beta^2})|-\frac{1}{2},\frac{1}{2}\rangle\nonumber\\
&&+(a_x +i a_y\sqrt{1-\beta^2})(b_x+ib_y\sqrt{1-\beta^2})|-\frac{1}{2},-\frac{1}{2}\rangle\},\label{abtri}\\
\hat{a}\otimes\hat{b}|-\frac{1}{2},-\frac{1}{2}\rangle&=&\frac{1}{\sqrt{[1+\beta^2(a_x^2-1)][1+\beta^2(b_x^2-1)]}}\{(a_x -i a_y\sqrt{1-\beta^2})
(b_x-ib_y\sqrt{1-\beta^2})|\frac{1}{2},\frac{1}{2}\rangle\nonumber\\
&&-\sqrt{1-\beta^2}b_z (a_x-ia_y\sqrt{1-\beta^2})|\frac{1}{2},-\frac{1}{2}\rangle\nonumber\\
&&-\sqrt{1-\beta^2}a_z (b_x-ib_y\sqrt{1-\beta^2})|-\frac{1}{2},\frac{1}{2}\rangle\nonumber\\
&&+(1-\beta^2)a_z b_z|-\frac{1}{2},-\frac{1}{2}\rangle\},\label{abtrii}\\
\hat{a}\otimes\hat{b}|\frac{1}{2},-\frac{1}{2}\rangle&=&\frac{1}{\sqrt{[1+\beta^2(a_x^2-1)][1+\beta^2(b_x^2-1)]}}
\{\sqrt{1-\beta^2}a_z (b_x-ib_y\sqrt{1-\beta^2})|\frac{1}{2},\frac{1}{2}\rangle\nonumber\\
&&-(1-\beta^2)a_z b_z|\frac{1}{2},-\frac{1}{2}\rangle\nonumber\\
&&+(a_x +i a_y\sqrt{1-\beta^2})(b_x-ib_y\sqrt{1-\beta^2})|-\frac{1}{2},\frac{1}{2}\rangle\nonumber\\
&&-\sqrt{1-\beta^2}b_z (a_x+ia_y\sqrt{1-\beta^2})|-\frac{1}{2},-\frac{1}{2}\rangle\},\label{abtriii}\\
\hat{a}\otimes\hat{b}|-\frac{1}{2},\frac{1}{2}\rangle&=&\frac{1}{\sqrt{[1+\beta^2(a_x^2-1)][1+\beta^2(b_x^2-1)]}}
\{\sqrt{1-\beta^2}b_z (a_x-ia_y\sqrt{1-\beta^2})|\frac{1}{2},\frac{1}{2}\rangle\nonumber\\
&&+(a_x -i a_y\sqrt{1-\beta^2})(b_x+ib_y\sqrt{1-\beta^2})|\frac{1}{2},-\frac{1}{2}\rangle\nonumber\\
&&-(1-\beta^2)a_z b_z|-\frac{1}{2},\frac{1}{2}\rangle\nonumber\\
&&-\sqrt{1-\beta^2}a_z (b_x+ib_y\sqrt{1-\beta^2})|-\frac{1}{2},-\frac{1}{2}\rangle\}\label{abtriv}
\end{eqnarray}
\end{mathletters}
for the boost in the $x$-direction. The calculation of $\langle\hat{a}\otimes\hat{b}\rangle$ is straightforward and is given by
\begin{eqnarray}
\langle\hat{a}\otimes\hat{b}\rangle&=&\frac{1}{\sqrt{[1+\beta^2(a_x^2-1)][1+\beta^2(b_x^2-1)]}}\{[a_x b_x+(1-\beta^2)a_z b_z]\cos(\Omega_{\vec{p}}+\Omega_{-\vec{p}})\nonumber\\
&&-(1-\beta^2)a_y b_y -\sqrt{1-\beta^2}(a_z b_x-b_z a_x)\sin(\Omega_{\vec{p}}+\Omega_{-\vec{p}})\}.
\label{expe}
\end{eqnarray}

It is interesting to note that in the ultra-relativistic limit, $\beta\to 1$, equation (\ref{expe}) becomes
\begin{equation}
\langle\hat{a}\otimes\hat{b}\rangle\rightarrow \frac{a_x}{|a_x|}\cdot\frac{b_x}{|b_x|}\cos(\Omega_{\vec{p}}+\Omega_{-\vec{p}}),
\end{equation}
implying that the joint measurements are not correlated at all. As a result, one might suspect that the entangled state satisfies
the Bell's inequality. We now consider the vectors $\vec{a}=(\frac{1}{\sqrt{2}},-\frac{1}{\sqrt{2}},0),
\vec{a}'=(-\frac{1}{\sqrt{2}},-\frac{1}{\sqrt{2}},0), \vec{b}=(0, 1, 0), \vec{b}'=(1, 0, 0)$ which lead to the maximum
violation of the Bell's inequality in the non-relativistic domain, $\Omega_{\vec{p}}=\Omega_{-\vec{p}}=0$ and $\beta=0$. Then the Bell observable
for the $4$ relevant joint measurements becomes
\begin{eqnarray}
&&\langle\hat{a}\otimes\hat{b}\rangle+\langle\hat{a}\otimes\hat{b}'\rangle+\langle\hat{a}'\otimes\hat{b}\rangle-\langle\hat{a}'\otimes\hat{b}'\rangle\nonumber\\
&&=\frac{2}{\sqrt{2-\beta^2}}(\sqrt{1-\beta^2}+\cos(\Omega_{\vec{p}}+\Omega_{-\vec{p}})).
\label{bells}
\end{eqnarray}
In the ultra-relativistic limit where $\beta=1$, the eq. (\ref{bells}) gives the maximum value of $2$ satisfying the Bell's inequality as expected.\\

Case II: $\Psi_{01}\rightarrow U(\Lambda)\Psi_{01}$\\
From eq. (\ref{ulpii}), we have
\begin{eqnarray}
U(\Lambda)\Psi_{01}=\frac{(\Lambda p)^0}{p^0}|\vec{p}_{\Lambda}, -\vec{p}_{\Lambda}\rangle\otimes[\frac{1}{\sqrt{2}}\cos\frac{\Omega_{\vec{p}}-\Omega_{-\vec{p}}}{2}(|\frac{1}{2},\frac{1}{2}\rangle-|-\frac{1}{2}, -\frac{1}{2}\rangle)\nonumber\\
+\frac{1}{\sqrt{2}}\sin\frac{\Omega_{\vec{p}}-\Omega_{-\vec{p}}}{2}(|\frac{1}{2},-\frac{1}{2}\rangle+|-\frac{1}{2}, \frac{1}{2}\rangle)].
\end{eqnarray}
From equations (\ref{abtri}) to (\ref{abtriv}), we obtain
\begin{eqnarray}
\langle\hat{a}\otimes\hat{b}\rangle&=&\frac{1}{\sqrt{[1+\beta^2(a_x^2-1)][1+\beta^2(b_x^2-1)]}}\{[-a_x b_x+(1-\beta^2)a_z b_z]\cos(\Omega_{\vec{p}}-\Omega_{-\vec{p}})\nonumber\\
&&+(1-\beta^2)a_y b_y +\sqrt{1-\beta^2}(a_z b_x+b_z a_x)\sin(\Omega_{\vec{p}}-\Omega_{-\vec{p}})\}.
\label{expei}
\end{eqnarray}
Then, in the ultra-relativistic limit, $\beta\to 1$, we have

\begin{equation}
\langle\hat{a}\otimes\hat{b}\rangle\rightarrow -\frac{a_x}{|a_x|}\cdot\frac{b_x}{|b_x|}\cos(\Omega_{\vec{p}}-\Omega_{-\vec{p}}),
\end{equation}
again, indicating the joint measurements, become uncorrelated in this limit. We consider the vectors
$\vec{a}=(-\frac{1}{\sqrt{2}},\frac{1}{\sqrt{2}},0),
\vec{a}'=(\frac{1}{\sqrt{2}},\frac{1}{\sqrt{2}},0), \vec{b}=(0, 1, 0), \vec{b}'=(1, 0, 0)$ which lead to the maximum
violation of the Bell's inequality in the non-relativistic regime.
Then the Bell observable for the $4$ relevant joint measurements becomes
\begin{eqnarray}
&&\langle\hat{a}\otimes\hat{b}\rangle+\langle\hat{a}\otimes\hat{b}'\rangle+\langle\hat{a}'\otimes\hat{b}\rangle-\langle\hat{a}'\otimes\hat{b}'\rangle\nonumber\\
&&=\frac{2}{\sqrt{2-\beta^2}}(\sqrt{1-\beta^2}+\cos(\Omega_{\vec{p}}-\Omega_{-\vec{p}})),
\label{bellsi}
\end{eqnarray}
thus giving same maximum value as in the case I. It can also be shown that one can obtain the same value for the Bell
observables given by eq. (\ref{bellsi}) for $U(\Lambda)\Psi_{10}$ and $U(\Lambda)\Psi_{11}$ implying eq. ({\ref{bellsi}) is the universal result.\\

\textbf{B: The momentum and the boost vectors not in the same plane.}

Case I: $\Psi_{00}\rightarrow U(\Lambda)\Psi_{00}$\\
From eq. (\ref{ulti}), we have

\begin{eqnarray}
U(\Lambda)\Psi_{00}&=&\frac{(\Lambda p)^0}{p^0}|\vec{p}_{\Lambda}, -\vec{p}_{\Lambda}\rangle\otimes\{(\cos\bar{\Omega}_{\vec{p}}\cos^{2}\eta+\cos\Delta\Omega_{\vec{p}}\sin^{2}\eta)\frac{1}{\sqrt{2}}(|\frac{1}{2},\frac{1}{2}\rangle+
|-\frac{1}{2}, -\frac{1}{2}\rangle)\nonumber\\
&&-\sin\bar{\Omega}_{\vec{p}}\cos\eta\frac{1}{\sqrt{2}}(|\frac{1}{2},-\frac{1}{2}\rangle-
|-\frac{1}{2}, \frac{1}{2}\rangle)\nonumber\\
&&+i\sin\Delta\Omega_{\vec{p}}\sin\eta\frac{1}{\sqrt{2}}(|\frac{1}{2},\frac{1}{2}\rangle-
|-\frac{1}{2}, -\frac{1}{2}\rangle)\nonumber\\
&&-i(-\cos\bar{\Omega}_{\vec{p}}+\cos\Delta\Omega_{\vec{p}})\sin\eta\cos\eta\frac{1}{\sqrt{2}}(|\frac{1}{2},-\frac{1}{2}\rangle+
|-\frac{1}{2}, \frac{1}{2}\rangle)\} ,
\end{eqnarray}

From equations (\ref{abtri}) to (\ref{abtriv}), we obtain
\begin{eqnarray}
\langle\hat{a}\otimes\hat{b}\rangle&=&\frac{1}{\sqrt{[1+\beta^2(a_x^2-1)][1+\beta^2(b_x^2-1)]}}\{~
\frac{\textsl{A}_++\textsl{A}_-}{2}\textsl{X}^{2}+\frac{\textsl{C}_++\textsl{C}_-}{2}\textsl{Y}^{2}\nonumber\\
&&+\frac{\textsl{E}_++\textsl{E}_-}{2}\textsl{Z}^{2}+\frac{\textsl{G}_++\textsl{G}_-}{2}\textsl{W}^{2}
-2[\frac{\textsl{B}_++\textsl{B}_-}{2}\textsl{XY}+\frac{\textsl{A}_+-\textsl{A}_-}{2i}\textsl{XZ}\nonumber\\
&&+\frac{\textsl{B}_+-\textsl{B}_-}{2i}\textsl{XW}-\frac{\textsl{D}_+-\textsl{D}_-}{2i}\textsl{YZ}
+\frac{\textsl{C}_+-\textsl{C}_-}{2i}\textsl{YW}+\frac{\textsl{F}_++\textsl{F}_-}{2}\textsl{ZW}~]~\}.
\label{expeii}
\end{eqnarray}

where
\begin{mathletters}
\begin{eqnarray}
\textsl{X}&=&\cos\bar{\Omega}_{\vec{p}}\cos^{2}\eta+\cos\Delta\Omega_{\vec{p}}\sin^{2}\eta,\label{x}\\
\textsl{Y}&=&\sin\bar{\Omega}_{\vec{p}}\cos\eta,\\
\textsl{Z}&=&\sin\Delta\Omega_{\vec{p}}\sin\eta,\\
\textsl{W}&=&-\cos\bar{\Omega}_{\vec{p}}\cos\eta\sin\eta+\cos\Delta\Omega_{\vec{p}}\sin\eta\cos\eta,\label{w}\\
\frac{\textsl{A}_++\textsl{A}_-}{2}&=&a_x b_x-(1-\beta^2)a_y b_y +(1-\beta^2)a_z b_z,\\
\frac{\textsl{C}_++\textsl{C}_-}{2}&=&-a_x b_x-(1-\beta^2)a_y b_y -(1-\beta^2)a_z b_z,\\
\frac{\textsl{E}_++\textsl{E}_-}{2}&=&-a_x b_x+(1-\beta^2)a_y b_y +(1-\beta^2)a_z b_z,\\
\frac{\textsl{G}_++\textsl{G}_-}{2}&=&a_x b_x+(1-\beta^2)a_y b_y -(1-\beta^2)a_z b_z,\\
\frac{\textsl{B}_++\textsl{B}_-}{2}&=&\frac{\textsl{D}_++\textsl{D}_-}{2}=\sqrt{1-\beta^2}(a_z b_x-b_z a_x),\\
\frac{\textsl{F}_++\textsl{F}_-}{2}&=&\frac{\textsl{H}_++\textsl{H}_-}{2}=\sqrt{1-\beta^2}(a_z b_x+b_z a_x),\\
\frac{\textsl{A}_+-\textsl{A}_-}{2i}&=&\frac{\textsl{E}_+-\textsl{E}_-}{2i}=\sqrt{1-\beta^2}(a_x b_y+b_x a_y),\\
\frac{\textsl{C}_+-\textsl{C}_-}{2i}&=&\frac{\textsl{G}_+-\textsl{G}_-}{2i}=\sqrt{1-\beta^2}(a_x b_y-b_x a_y),\\
\frac{\textsl{B}_+-\textsl{B}_-}{2i}&=&\frac{\textsl{H}_+-\textsl{H}_-}{2i}=(1-\beta^2)(a_z b_y+b_z a_y),\\
\frac{\textsl{D}_+-\textsl{D}_-}{2i}&=&\frac{\textsl{F}_+-\textsl{F}_-}{2i}=(1-\beta^2)(a_z b_y-b_z a_y)
\end{eqnarray}
\end{mathletters}

It is interesting to note that in the ultra-relativistic limit, $\beta\to 1$, equation (\ref{expeii}) becomes
\begin{equation}
\langle\hat{a}\otimes\hat{b}\rangle\rightarrow \frac{a_x}{|a_x|}\cdot\frac{b_x}{|b_x|}(\textsl{X}^{2}-\textsl{Y}^{2}-\textsl{Z}^{2}+\textsl{W}^{2}),
\end{equation}
implying that the joint measurements are not correlated at all. As a result, one might suspect that the entangled state satisfies
the Bell's inequality. We now consider the vectors $\vec{a}=(\frac{1}{\sqrt{2}},-\frac{1}{\sqrt{2}},0),
\vec{a}'=(-\frac{1}{\sqrt{2}},-\frac{1}{\sqrt{2}},0), \vec{b}=(0, 1, 0), \vec{b}'=(1, 0, 0)$ which lead to the maximum
violation of the Bell's inequality in the non-relativistic domain, $\Omega_{\vec{p}}=\Omega_{-\vec{p}}=0$ and $\beta=0$. Then the Bell observable
for the $4$ relevant joint measurements becomes
\begin{eqnarray}
&&\langle\hat{a}\otimes\hat{b}\rangle+\langle\hat{a}\otimes\hat{b}'\rangle+\langle\hat{a}'\otimes\hat{b}\rangle-\langle\hat{a}'\otimes\hat{b}'\rangle\nonumber\\
&&=\frac{2}{\sqrt{2-\beta^2}}\{(\textsl{X}^{2}-\textsl{Y}^{2}-\textsl{Z}^{2}+\textsl{W}^{2})+(\textsl{X}^{2}+\textsl{Y}^{2}-\textsl{Z}^{2}-\textsl{W}^{2})\sqrt{1-\beta^2}\}.
\label{bellss}
\end{eqnarray}
In the ultra-relativistic limit where $\beta=1$, the eq. (\ref{bellss}) gives the maximum value of $2$ satisfying the Bell's inequality as expected.(Appendix)\\

Case II: $\Psi_{01}\rightarrow U(\Lambda)\Psi_{01}$\\
From eq. (\ref{ultii}), we have

\begin{eqnarray}
U(\Lambda)\Psi_{01}&=&\frac{(\Lambda p)^0}{p^0}|\vec{p}_{\Lambda}, -\vec{p}_{\Lambda}\rangle\otimes\{\cos\Delta\Omega_{\vec{p}}\frac{1}{\sqrt{2}}(|\frac{1}{2},\frac{1}{2}\rangle-
|-\frac{1}{2}, -\frac{1}{2}\rangle)\nonumber\\
&&+\sin\Delta\Omega_{\vec{p}}\cos\eta\frac{1}{\sqrt{2}}(|\frac{1}{2},-\frac{1}{2}\rangle+
|-\frac{1}{2}, \frac{1}{2}\rangle)\nonumber\\
&&+i\sin\Delta\Omega_{\vec{p}}\sin\eta\frac{1}{\sqrt{2}}(|\frac{1}{2},\frac{1}{2}\rangle+
|-\frac{1}{2}, -\frac{1}{2}\rangle)\},\label{ulpiip}
\end{eqnarray}

From equations (\ref{abtri}) to (\ref{abtriv}), we obtain
\begin{eqnarray}
\langle\hat{a}\otimes\hat{b}\rangle&=&\frac{1}{\sqrt{[1+\beta^2(a_x^2-1)][1+\beta^2(b_x^2-1)]}}\{~
\frac{\textsl{G}_++\textsl{G}_-}{2}\textsl{X}~'^{2}+\frac{\textsl{A}_++\textsl{A}_-}{2}\textsl{Y}~'^{2}\nonumber\\
&&+\frac{\textsl{E}_++\textsl{E}_-}{2}\textsl{Z}~'^{2}+2[\frac{\textsl{F}_++\textsl{F}_-}{2}\textsl{X}~'\textsl{Z}~'
+\frac{\textsl{A}_+-\textsl{A}_-}{2i}\textsl{Y}~'\textsl{Z}~'-\frac{\textsl{B}_+-\textsl{B}_-}{2i}\textsl{X}~'\textsl{Y}~'~]~\}.
\label{expeiii}
\end{eqnarray}

where
\begin{mathletters}
\begin{eqnarray}
\textsl{X}~'&=&\sin\Delta\Omega_{\vec{p}}\cos\eta,\\
\textsl{Y}~'&=&\sin\Delta\Omega_{\vec{p}}\sin\eta,\\
\textsl{Z}~'&=&\cos\Delta\Omega_{\vec{p}}
\end{eqnarray}
\end{mathletters}

Then, in the ultra-relativistic limit, $\beta\to 1$, we have

\begin{equation}
\langle\hat{a}\otimes\hat{b}\rangle\rightarrow -\frac{a_x}{|a_x|}\cdot\frac{b_x}{|b_x|}(-\textsl{X}~'^{2}-\textsl{Y}~'^{2}+\textsl{Z}~'^{2})
=-\frac{a_x}{|a_x|}\cdot\frac{b_x}{|b_x|}\cos2\Delta\Omega_{\vec{p}},
\end{equation}
again, indicating the joint measurements, become uncorrelated in this limit. We consider the vectors
$\vec{a}=(-\frac{1}{\sqrt{2}},\frac{1}{\sqrt{2}},0),
\vec{a}'=(\frac{1}{\sqrt{2}},\frac{1}{\sqrt{2}},0), \vec{b}=(0, 1, 0), \vec{b}'=(1, 0, 0)$ which lead to the maximum
violation of the Bell's inequality in the non-relativistic regime.
Then the Bell observable for the $4$ relevant joint measurements becomes
\begin{eqnarray}
&&\langle\hat{a}\otimes\hat{b}\rangle+\langle\hat{a}\otimes\hat{b}'\rangle+\langle\hat{a}'\otimes\hat{b}\rangle-\langle\hat{a}'\otimes\hat{b}'\rangle\nonumber\\
&&=\frac{2}{\sqrt{2-\beta^2}}\{(-\textsl{X}~'^{2}-\textsl{Y}~'^{2}+\textsl{Z}~'^{2})+(\textsl{X}~'^{2}-\textsl{Y}~'^{2}+\textsl{Z}~'^{2})\sqrt{1-\beta^2}\}\nonumber\\
&=&\frac{2}{\sqrt{2-\beta^2}}\{\cos2\Delta\Omega_{\vec{p}}+(\cos^{2}\eta+\sin^{2}\eta\cos2\Delta\Omega_{\vec{p}})\sqrt{1-\beta^2}\}
\label{bellsii}
\end{eqnarray}
thus giving same maximum value as in the case I. It can also be shown that one can obtain the same value for the Bell
observables given by eq. ({\ref{bellsii}) for $U(\Lambda)\Psi_{10}$ and $U(\Lambda)\Psi_{11}$ implying eq. ({\ref{bellsii}) is the universal result.

These agree with our previous results\cite{1} which didn't take
into account the general momentum. It can also be shown that
similar results would be obtained for the case\cite{20} in which
one observer is in the rest frame and the other observer is in the
moving frame and do the joint measurements of spins. Now, one can
see that the quantum correlation approaches to the classical
correlation when the speed of the moving observer reaches the
speed of light and in both cases, the information in the vertical
direction to the boost axis is lost. This is somewhat analogous to
the cases of $\beta$-decay of nuclei and high energy electrons and
positrons emitted in the decay of muons for which emitted
electrons and positrons are polarized such that their spins tend
to lie in the same direction of the motion and their projections
of the spins in the direction of the motion became $\pm1$ for the
relativistic particles\cite{21}. It should be noted that if one
simply rotates the spin directions instead of using the
relativistic spin observables, then the entanglement between the
spins of the Bell states are not changed and the results of the
spin measurements would be exactly same as if they were done in
the rest frame thus give the maximum violation of the Bell
inequality. It is interesting to note that the entanglement is
still remained though it is degraded, when Bell's inequality is
satisfied.  The most plausible reason for this is that the quantum
correlations in the vertical direction to the boost are lost  and
become classical.  So we can also conclude that the Bell's
inequality is not always violated for entangled state in special
relativity.

\section{summary}
In this work, we studied the Lorentz transformed entangled Bell
states and the Bell observables in the case of general momentum to
investigate whether the Bell's inequality is always violated in
special relativity. We have calculated the Bell observable for the
joint $4$ measurements and found that the results are universal
for all entangled states:
\begin{eqnarray}
c(\vec{a},\vec{a}',\vec{b},\vec{b}')&=&\langle\hat{a}\otimes\hat{b}\rangle+\langle\hat{a}\otimes\hat{b}'\rangle+\langle\hat{a}'\otimes\hat{b}\rangle
-\langle\hat{a}'\otimes\hat{b}'\rangle\nonumber\\
&=&\frac{2}{\sqrt{2-\beta^2}}(1+\sqrt{1-\beta^2}),
\nonumber
\end{eqnarray}
where $\hat{a}, \hat{b}$ are the relativistic spin observables
derived from the Pauli-Lubanski pseudo vector. It turn out that
the Bell observable is a monotonically decreasing function of
$\beta$ and approaches the limit value of $2$ when $\beta=1$
indicating that the Bell's inequality is not always violated in
the ultra-relativistic limit. It is also shown that quantum
information, along the perpendicular direction to the boost, is
eventually lost and Bell's inequality is not always violated for
entangled states in special relativity. This could impose
restrictions to certain quantum information processing such as
quantum cryptography using massive particles.  Unless both sender
and receiver measures along the boost direction, there will be
information loss.\vspace{2.0cm}

\centerline{\bf Acknowledgements}

This work was supported by the Korean Ministry of Science and
Technology through the Creative Research Initiatives Program under
Contact No. M1-0116-00-0008. We are also indebted to M. Czachor for valuable discussions.

\appendix

\section*{ Derivation of eq.(39)}
We have already known as follows,
$\cosh\alpha=\frac{1}{\sqrt{1-\beta^2}},\cosh\delta=\frac{p^0}{m}$ ~from (\ref{rotai}) and (\ref{rotaii}),\\
and defined as follows,
$\cos\eta=\frac{\cos\theta}{r},\sin\eta=\frac{\sin\theta\sin\phi}{r},r=\sqrt{\sin^2 \theta\sin^2 \phi +\cos^2 \theta}$ ,
and $\bar{\Omega}_{\vec{p}}=\frac{\Omega_{\vec{p}}+\Omega_{-\vec{p}}}{2},\Delta\Omega_{\vec{p}}=\frac{\Omega_{\vec{p}}-\Omega_{-\vec{p}}}{2}$.

From Eq.(\ref{ome21}) and (\ref{ome22}),
we obtain
\begin{eqnarray}
\cos\bar{\Omega}_{\vec{p}}
&=&\cos\frac{\Omega_{\hat{p}}}{2}\cos\frac{\Omega_{-\hat{p}}}{2}-\sin\frac{\Omega_{\hat{p}}}{2}\sin\frac{\Omega_{-\hat{p}}}{2}\nonumber\\
&=&\frac{\cosh^{2}\frac{\alpha}{2}\cosh^{2}\frac{\delta}{2}-\sinh^{2}\frac{\alpha}{2}\sinh^{2}\frac{\delta}{2}}
{[(\frac{1}{2}+\frac{1}{2}\cosh\alpha\cosh\delta)^{2}-(\frac{1}{2}\sinh\alpha\sinh\delta\sin\theta\cos\phi)^{2}]^{\frac{1}{2}}}\nonumber\\
&=&\frac{\frac{\cosh^{2}\frac{\alpha}{2}\cosh^{2}\frac{\delta}{2}}{\sinh^{2}\frac{\alpha}{2}\sinh^{2}\frac{\delta}{2}}-1}
{[(\frac{1+\cosh\alpha\cosh\delta}{2\sinh^{2}\frac{\alpha}{2}\sinh^{2}\frac{\delta}{2}})^{2}-(\frac{\sinh\alpha\sinh\delta}{2\sinh^{2}\frac{\alpha}{2}\sinh^{2}\frac{\delta}{2}})^{2}(1-r^2)]^{\frac{1}{2}}}\nonumber\\
&=&\frac{\coth^{2}\frac{\alpha}{2}\coth^{2}\frac{\delta}{2}-1}{[(\coth^{2}\frac{\alpha}{2}\coth^{2}\frac{\delta}{2}+1)^{2}-4\coth^{2}\frac{\alpha}{2}\coth^{2}\frac{\delta}{2}(1-r^{2})]^{\frac{1}{2}}}\nonumber\\
&=&\frac{\coth^{2}\frac{\alpha}{2}\coth^{2}\frac{\delta}{2}-1}{[(\coth^{2}\frac{\alpha}{2}\coth^{2}\frac{\delta}{2}-1)^{2}+4\coth^{2}\frac{\alpha}{2}\coth^{2}\frac{\delta}{2}r^{2}]^{\frac{1}{2}}}\nonumber\\
&=&\frac{t-1}{[(t-1)^2+4tr^2]^{\frac{1}{2}}},\label{app1}
\end{eqnarray}
where $t=\coth^{2}\frac{\alpha}{2}\coth^{2}\frac{\delta}{2}$,$~~1\leq t~$,\\
and
\begin{eqnarray}
\cos\Delta\Omega_{\vec{p}}
&=&\cos\frac{\Omega_{\hat{p}}}{2}\cos\frac{\Omega_{-\hat{p}}}{2}+\sin\frac{\Omega_{\hat{p}}}{2}\sin\frac{\Omega_{-\hat{p}}}{2}\nonumber\\
&=&\frac{\cosh^{2}\frac{\alpha}{2}\cosh^{2}\frac{\delta}{2}-\sinh^{2}\frac{\alpha}{2}\sinh^{2}\frac{\delta}{2}(\sin^{2}\theta\cos^{2}\phi-\sin^{2}\theta\sin^{2}\phi-\cos^{2}\theta)}
{[(\frac{1}{2}+\frac{1}{2}\cosh\alpha\cosh\delta)^{2}-(\frac{1}{2}\sinh\alpha\sinh\delta\sin\theta\cos\phi)^{2}]^{\frac{1}{2}}}\nonumber\\
&=&\frac{\frac{\cosh^{2}\frac{\alpha}{2}\cosh^{2}\frac{\delta}{2}}{\sinh^{2}\frac{\alpha}{2}\sinh^{2}\frac{\delta}{2}}-(1-2r^{2})}
{[(\frac{1+\cosh\alpha\cosh\delta}{2\sinh^{2}\frac{\alpha}{2}\sinh^{2}\frac{\delta}{2}})^{2}-(\frac{\sinh\alpha\sinh\delta}{2\sinh^{2}\frac{\alpha}{2}\sinh^{2}\frac{\delta}{2}})^{2}(1-r^2)]^{\frac{1}{2}}}\nonumber\\
&=&\frac{\coth^{2}\frac{\alpha}{2}\coth^{2}\frac{\delta}{2}-(1-2r^{2})}{[(\coth^{2}\frac{\alpha}{2}\coth^{2}\frac{\delta}{2}+1)^{2}-4\coth^{2}\frac{\alpha}{2}\coth^{2}\frac{\delta}{2}(1-r^{2})]^{\frac{1}{2}}}\nonumber\\
&=&\frac{(\coth^{2}\frac{\alpha}{2}\coth^{2}\frac{\delta}{2}-1)+2r^{2}}{[(\coth^{2}\frac{\alpha}{2}\coth^{2}\frac{\delta}{2}-1)^{2}+4\coth^{2}\frac{\alpha}{2}\coth^{2}\frac{\delta}{2}r^{2}]^{\frac{1}{2}}}\nonumber\\
&=&\frac{(t-1)+2r^{2}}{[(t-1)^2+4tr^2]^{\frac{1}{2}}}.\label{app2}
\end{eqnarray}
From (\ref{x}) to (\ref{w}), and from (\ref{app1}) and (\ref{app2}), we get
\begin{eqnarray}
\textsl{X}^{2}-\textsl{Y}^{2}-\textsl{Z}^{2}+\textsl{W}^{2}
&=&\cos^{2}\bar{\Omega}_{\hat{p}}\cos^{2}\eta+\cos^{2}\Delta\Omega_{\hat{p}}\sin^{2}\eta-\sin^{2}\bar{\Omega}_{\hat{p}}\cos^{2}\eta-\sin^{2}\Delta\Omega_{\hat{p}}\sin^{2}\eta \nonumber\\
&=&2(\cos^{2}\bar{\Omega}_{\hat{p}}\cos^{2}\eta+\cos^{2}\Delta\Omega_{\hat{p}}\sin^{2}\eta)-1 \nonumber\\
&=&2\frac{(t-1)^{2}\cos^{2}\eta+(t-1+2r^{2})^{2}\sin^{2}\eta}{(t-1)^{2}+4tr^{2}}-1 \nonumber\\
&=&1-8r^{2}\cos^{2}\eta\frac{t+(1-r^{2})\tan^{2}\eta}{(t-1)^{2}+4tr^{2}},\label{app3}
\end{eqnarray}
and
\begin{eqnarray}
\textsl{X}^{2}+\textsl{Y}^{2}-\textsl{Z}^{2}-\textsl{W}^{2}
&=&\cos^{2}\bar{\Omega}_{\hat{p}}\cos^{2}\eta\cos2\eta-\cos^{2}\Delta\Omega_{\hat{p}}\sin^{2}\eta\cos2\eta+\sin^{2}\bar{\Omega}_{\hat{p}}\cos^{2}\eta \nonumber\\
&&-\sin^{2}\Delta\Omega_{\hat{p}}\sin^{2}\eta+4\cos\bar{\Omega}_{\hat{p}}\cos\Delta\Omega_{\hat{p}}\sin^{2}\eta\cos^{2}\eta \nonumber\\
&=&\cos2\eta+(1-\cos2\eta)\{\cos^{2}\Delta\Omega_{\hat{p}}-\cos^{2}\eta(\cos\bar{\Omega}_{\hat{p}}-\cos\Delta\Omega_{\hat{p}})^{2}\} \nonumber\\
&=&\cos2\eta+(1-\cos2\eta)\{\frac{(t-1+2r^{2})^{2}-\cos^{2}\eta(-2r^{2})^{2}}{(t-1)^{2}+4tr^{2}}\} \nonumber\\
&=&1-8r^{2}\sin^{2}\eta\frac{1-r^{2}\sin^{2}\eta}{(t-1)^{2}+4tr^{2}}.\label{app4}
\end{eqnarray}
From (\ref{app3}), we define
\begin{eqnarray}
f(t)&=&\frac{t+a}{(t-1)^{2}+4r^{2}t},\label{app5}\\
\textmd{and then}\nonumber\\
\frac{df(t)}{dt}&=&-\frac{(t-1)(t+2a+1)+4r^{2}a}{\{ (t-1)^{2}+4r^{2}t\}^{2}}<0,~\textmd{for} ~t\geq1,~\forall\theta ~\textmd{and}~\forall\phi,\label{app6}
\end{eqnarray}
where ~$a=(1-r^{2})\tan^{2}\eta\geq 0$.\\
From (\ref{app5}) and (\ref{app6}), we obtain
\begin{equation}
0=f(\infty)\leq f(t)\leq f(1)=\frac{1+a}{4r^{2}}
\end{equation}
and
\begin{eqnarray}
1-8r^{2}\cos^{2}\eta f(1)
&=&1-8r^{2}\cos^{2}\eta\frac{1+(1-r^{2})\tan^{2}\eta}{4r^{2}}\nonumber\\
&=&2\sin^{2}\theta\sin^{2}\phi-1,
\end{eqnarray}
therefore
\begin{equation}
2\sin^{2}\theta\sin^{2}\phi-1\leq \textsl{X}^{2}-\textsl{Y}^{2}-\textsl{Z}^{2}+\textsl{W}^{2} \leq 1.
\end{equation}

From (\ref{app4}), we define
\begin{eqnarray}
g(t)&=&\frac{b}{(t-1)^{2}+4r^{2}t},\label{app10}\\
\textmd{and then}\nonumber\\
\frac{dg(t)}{dt}&=&-2b\frac{(t-1)+2r^{2}}{\{ (t-1)^{2}+4r^{2}t\}^{2}}\leq 0,~\textmd{for} ~t\geq1,~\forall\theta ~\textmd{and}~\forall\phi,\label{app11}
\end{eqnarray}
where ~$b=1-r^{2}\sin^{2}\eta\geq 0$. \\
From (\ref{app10}) and (\ref{app11}), we have
\begin{equation}
0=g(\infty)\leq g(t)\leq g(1)=\frac{b}{4r^{2}}
\end{equation}
and
\begin{eqnarray}
1-8r^{2}\sin^{2}\eta g(1)
&=&1-8r^{2}\sin^{2}\eta\frac{1-r^{2}\sin^{2}\eta}{4r^{2}} \nonumber\\
&=&1-2\sin^{2}\eta(1-r^{2}\sin^{2}\eta) \nonumber\\
&=&\cos2\eta+\frac{r^2}{2}(1-\cos2\eta)^{2} \nonumber\\
&\geq &\cos2\eta,
\end{eqnarray}
therefore
\begin{equation}
\cos2\eta\leq \textsl{X}^{2}+\textsl{Y}^{2}-\textsl{Z}^{2}-\textsl{W}^{2} \leq 1.
\end{equation}

\end{document}